\documentclass[final]{ustcstep}

\usepackage{lineno}
\usepackage{graphicx}
\usepackage{lscape}
\usepackage{multirow}
\usepackage{color}
\usepackage{url}

\newcommand\addr[2]{{\footnotesize \it $^{#1}$#2}\\}
\usepackage[pdfborder={0 0 0},urlcolor=blue,breaklinks]{hyperref}

\usepackage{enumerate}
\usepackage{natbib}

\begin{document}

\title{Full Halo Coronal Mass Ejections: Arrival at the Earth}

\author{Chenglong Shen$^{1,2}$,Yuming Wang$^{1}$, Zonghao Pan$^{1}$, Bin Miao$^{1}$, Pinzhong Ye$^{1}$, S. Wang$^{1}$\\
	 \addr{}{$^{1}$CAS Key Laboratory of Geospace Environment,
        Department of Geophysics and Planetary Sciences,}\\
        \addr{}{$^{2}$University of Science and Technology of China, Hefei, Anhui 230026, China}\\
    \addr{}{\href{mailto:clshen@ustc.edu.cn}{clshen@ustc.edu.cn}}
}

\maketitle
\tableofcontents

\begin{abstract}
A geomagnetic storm is mainly caused by a front-side coronal mass ejection (CME) hitting the Earth and then interacting with the magnetosphere. However, not all front-side CMEs can hit the Earth. Thus, which CMEs hit the Earth and when they do so are important issues in the study and forecasting of space weather. In our previous work (Shen et al., 2013), the de-projected parameters of the full-halo coronal mass ejections (FHCMEs) that occurred from 2007 March 1 to 2012 May 31 were estimated, and there are 39 front-side events could be fitted by the GCS model. In this work, we continue to study whether and when these front-side FHCMEs (FFHCMEs) hit the Earth. It is found that 59\% of these FFHCMEs hit the Earth, and for central events, whose deviation angles $\epsilon$, which are the angles between the propagation direction and the Sun-Earth line, are smaller than 45 degrees, the fraction increases to 75\%. After checking the  deprojected angular widths of the CMEs, we found that all of the Earth-encountered CMEs satisfy a simple criterion that the angular width ($\omega$) is larger than twice the deviation angle ($\epsilon$). This result suggests that some simple criteria can be used to forecast whether a CME could hit the Earth. Furthermore, for Earth-encountered CMEs, the transit time is found to be roughly anti-correlated with the de-projected velocity, but some events significantly deviate from the linearity. For CMEs with similar velocities, the differences of their transit times can be up to several days. Such deviation is further demonstrated to be mainly caused by the CME geometry and propagation direction, which are essential in the forecasting of CME arrival.
\end{abstract}

\section{Introduction}

The halo coronal mass ejections (CMEs), which appear to surround the occulting disk of coronagraphs, are preliminarily supposed to be propagating along the Sun-Earth line\citep{Howard:1982vo}.
Under this assumption, the front-side halo CMEs might be good candidates for Earth-impacted CMEs; however, not all of the front-side halo CMEs can hit the Earth. The ratio of the front-side halo CMEs hitting the Earth varied from 65\% to 80\%, which has been reported in different literature reports\citep[][and reference therein]{Yermolaev:2006ic}. 
In addition, most works are concerned about the geoeffectiveness of halo CMEs\citep[e.g.][]{webb2002,Wang:2002ki,Zhao:2003ds,zhang2007b,Gopalswamy2007}. 
The ratio of the front-side halo CMEs with geoeffectiveness varied from 45\% to 71\%. 
All of these works suggested that not all front-side halo CMEs can hit the Earth. 
Thus, what type of front-side halo CMEs can hit the Earth has been discussed by many authors\citep[e.g.][]{Wang:2002ki,2003ApJ...582..520Z,Kim:2005gm,Moon2005}. 
Before the launch of the Solar Terrestrial Relations Observatory (STEREO), only coronagraph images and in situ measurements could be used to observe the CMEs near the Sun and the interplanetary CMEs (ICME) near the Earth respectively. 
Thus, direct connections between the CMEs near the Sun and the ICMEs near 1 AU might be unclear, especially during the solar maximum. Recently, using the large field of view observations from the Heliospheric imagers in the Sun-Earth-Connection Coronal and Heliospheric Investigation (SECCHI) \citep{Howard2008a} onboard STEREO, 
the propagation of CMEs could be well tracked continuously from the Sun to 1 AU. In this manner, the ejecta observed near the Earth and the CMEs that occurred near the Sun can be related in a more precise way. Can we re-investigate how many and what type of front-side halo CMEs can hit the Earth? In addition, we note that the apparent angular width threshold used to define the halo CMEs in previous works varied greatly, such as 120$^\circ$, 130$^\circ$, 140$^\circ$ and 360$^\circ$. If we apply the apparent angular width = 360$^\circ$ only, will the ratio and the criteria of the front-side full halo CMEs arriving at the Earth be changed?

In addition, if a CME can hit the Earth, its arrival time becomes an important issue in space weather forecasting. Recently, various kinematics and magnetohydrodynamic (MHD) models have been developed to forecast the arrival time of CMEs\cite[e.g.][and reference therein]{Fry:2003vh,Odstrcil2004,Toth2005,MckennaLawlor:2006gm,2006SoPh..238..167F,shen:2007wwa,Feng:2007tl,Shen:2010eb,Feng2010}
In those models, the velocity, propagation direction and angular width of the CMEs are used as the initial parameters. However, the following basic questions are still not fully answered. 
What are the key parameters that determine the transit time of the CMEs from the Sun to the Earth? 
What is the extent of influence of the leading parameter? 
The CME's initial speed has been correlated with the transit time of the CME from the Sun to 1 AU\citep{Cane2000,Wang:2002ki,2003ApJ...582..520Z,Schwenn2005,Shanmugaraju2014}. 
Some simple equations were established to calculate the possible arrival time of the CMEs based on their initial velocities. 
However, the deviation between the calculated transit times and the observations is large. 
One possible reason for this deviation is that the CME's velocity might change greatly during its propagation in the interplanetary space due to the influence of the background solar wind\citep[e.g.][]{Gopalswamy2000}.
Using a constant acceleration (or deceleration) assumption, \cite{2001JGR...10629207G,Gopalswamy2005a} 
developed an empirical CME arrival (ECA) model to predict the arrival time of Earth-directed CMEs. However, the acceleration (or deceleration) may also be changed during the propagation of CMEs in the interplanetary space. Recently, some other CME propagation time forecasting models were developed based on aerodynamic drag models\citep[e.g.][]{Vrsnak2012}.
The aerodynamic drag models assume that the acceleration (or deceleration) of CMEs depends on the velocity difference between the CME and the background solar wind. 
Most of the above works are mainly focused on the velocity of the CME.  
Are there any other parameters that would exert significant influence on the propagation time of CMEs from the Sun to the Earth? How significant the influence of these parameters? The propagation direction of CMEs might be another important parameter. It has been taken into account in many CME arrival time forecasting models, such as the advanced version of the drag-based model (DBM, http://oh.geof.unizg.hr/DBM/dbm.php), the ENLIL model and other MHD models\citep[e. g.][]{Odstrcil2003}. Recently, based on a self-similar expansion assumption and a theoretical computation\citep{Davies2012},  \citet{Mostl2012a} suggested that the propagation time of a CME is influenced by its propagation direction and the angular width. 

In our previous work (\citet{Shen2013d}, referred to as Paper I hereafter), the projection effect of full-halo CMEs (FHCMEs) listed in the CDAW CME catalog\citep{2004JGRA..10907105Y} 
with an apparent angular width of 360$^\circ$ occurred from 2007 March 1 to 2012 May 31 was studied. In Paper I, the Graduated Cylindrical Shell (GCS) model
\citep{Thernisien2006,2009SoPh..256..111T,Thernisien:2011jy} was applied on the STEREO/COR2 and SOHO observations to obtain the de-projected kinematic parameters of these FHCMEs. Among the total of 88 events studied in paper I, 48 events originated from the front of the solar disk. Table \ref{list} in this paper and Table B in our online list (http://space.ustc.edu.cn/dreams/fhcmes/) present the parameters of these front-side full halo CMEs (FFHCMEs). Of the total of 48 FFHCMEs, there are nine events that could not be fitted by the GCS model. The remaining 39 events with well-established de-projected parameters will be studied in detail here. In this work, we will verify whether these FFHCMEs hit the Earth by using the continuous CME propagation observations of COR2, HI1, HI2 on board the STEREO spacecraft and the in situ measurements of the WIND and ACE satellites. Next, we attempt to answer the main questions of which and when the FFHCMEs will hit the Earth based on the de-projected parameters. In section 2, we introduce the method to determine the interplanetary counterpart for a given FFHCME. Based on the list of the FFHCMEs and their associated interplanetary counterparts, the type of the Earth-encountered FFHCMEs will be discussed in section 3. In section 4, the parameters that affect the transit time of CMEs from the Sun to the Earth will be discussed. A conclusion and some discussions of the results will be provided in the last section.

\section{Methods}

\begin{table*}
\caption{The GCS model's parameters and the times of the associated ICMEs of the FFHCMEs occurred from 2007 to 2012 May 31}
\begin{tabular}{|c|c|c|c|c|c|c|c|c|}
\hline No & CME date & Direction & $\epsilon$ & $\omega$ & $V_{GCS}$ & $T_{Shock}$ & $T_{ICME}$ begin& $T_{ICME}$ end\\
\hline  1&2009/12/16 04:30:03&E07,N09& 11& 45& 411&---&2009/12/19 09:49&2009/12/20 09:22\\
\hline  2&2010/02/07 03:54:03&E06,S07&  9& 81& 481&2010/02/11 00:00&2010/02/11 13:00&2010/02/11 22:00\\
\hline  3&2010/02/12 13:42:04&E01,N11& 11& 84& 550&2010/02/15 17:40&2010/02/16 04:00&2010/02/16 12:00\\
\hline  4&2010/04/03 10:33:58&E01,S27& 27& 84& 853&2010/04/05 07:56&2010/04/05 12:00&2010/04/06 16:00\\
\hline  5&2010/05/23 18:06:05&W16,N07& 17& 70& 365&2010/05/28 01:58&2010/05/28 19:00&2010/05/29 17:00\\
\hline  6&2010/05/24 14:06:05&W26,S06& 26& 63& 552&---&---&---\\
\hline  7&2010/08/01 13:42:05&E38,N20& 42& 93&1262&2010/08/03 17:00&2010/08/04 10:00&2010/08/05 02:00\\
\hline  8&2010/08/07 18:36:06&E36,S06& 36& 83& 779&---&2010/08/11 05:00&2010/08/12 17:00\\
\hline  9&2010/08/14 10:12:05&W42,S11& 43&119& 864&---&---&---\\
\hline 10&2010/12/14 15:36:05&W35,N39& 50&112& 856&---&---&---\\
\hline 11&2011/02/14 18:24:05&W08,N01&  8& 61& 365&---&2011/02/18 10:30&2011/02/18 19:30\\
\hline 12&2011/02/15 02:24:05&W05,S07&  8&140& 764&2011/02/18 01:00&2011/02/18 20:00&2011/02/20 08:00\\
\hline 13&2011/03/07 20:00:05&W34,N33& 45&104&1933&---&---&---\\
\hline 14&2011/06/02 08:12:06&E30,S03& 30& 92& 961&2011/06/04 20:00&2011/06/05 02:00&2011/06/05 18:00\\
\hline 15&2011/06/07 06:49:12&  --- &---&---&----&---&---&---\\
\hline 16&2011/06/21 03:16:10&E20,N07& 21& 93& 964&2011/06/23 02:00&2011/06/23 06:00&2011/06/24 06:00\\
\hline 17&2011/08/03 14:00:07&W10,N12& 15&124& 925&2011/08/04 21:15&2011/08/05 03:30&2011/08/05 17:30\\
\hline 18&2011/08/04 04:12:05&W36,N24& 42&107&----&2011/08/05 17:30&2011/08/06 22:00&2011/08/07 22:00\\
\hline 19&2011/08/09 08:12:06&W45,N16& 47&133&1594&---&---&---\\
\hline 20&2011/09/06 02:24:05&  --- &---&---&----&---&2011/09/08 10:00&2011/09/09 12:00\\
\hline 21&2011/09/06 23:05:57&W41,N19& 44&116& 901&2011/09/09 12:00&2011/09/10 03:00&2011/09/10 15:00\\
\hline 22&2011/09/22 10:48:06&E72,N06& 72&131&1823&---&---&---\\
\hline 23&2011/09/24 12:48:07&E47,N06& 47&119&1768&2011/09/26 12:00&2011/09/26 20:00&2011/09/28 00:00\\
\hline 24&2011/09/24 19:36:06&  --- &---&---&----&---&---&---\\
\hline 25&2011/10/22 01:25:53&  --- &---&---&----&---&---&---\\
\hline 26&2011/10/22 10:24:05&  --- &---&---&----&2011/10/24 17:38&2011/10/25 00:00&2011/10/25 16:00\\
\hline 27&2011/10/27 12:00:06&E42,N26& 48& 51&----&---&---&---\\
\hline 28&2011/11/09 13:36:05&E36,N24& 42&172&1074&2011/11/12 05:26&2011/11/12 14:51&2011/11/13 11:09\\
\hline 29&2011/11/26 07:12:06&W35,N17& 38&177& 900&2011/11/28 20:51&2011/11/29 00:12&2011/11/29 04:53\\
\hline 30&2012/01/02 15:12:40&  --- &---&---&----&2012/01/05 15:50&2012/01/04 00:00&2012/01/06 02:43\\
\hline 31&2012/01/16 03:12:10&E57,N39& 64&124& 958&---&---&---\\
\hline 32&2012/01/19 14:36:05&E17,N43& 45&141&1090&2012/01/22 05:10&2012/01/23 00:13&2012/01/24 15:09\\
\hline 33&2012/01/23 04:00:05&W16,N41& 43&193&1906&2012/01/24 14:30&---&---\\
\hline 34&2012/01/26 04:36:05&W71,N56& 79& 85&1033&---&---&---\\
\hline 35&2012/01/27 18:27:52&W78,N27& 79&179&1807&2012/01/30 15:56&---&---\\
\hline 36&2012/02/09 21:17:36&E42,N29& 49& 79& 648&---&---&---\\
\hline 37&2012/02/10 20:00:05&E25,N20& 31& 74& 583&---&2012/02/14 17:58&2012/02/16 05:33\\
\hline 38&2012/02/23 08:12:06&W61,N28& 64&135& 442&2012/02/26 20:58&2012/02/27 17:53&2012/02/28 15:40\\
\hline 39&2012/03/04 11:00:07&E41,N27& 47&150&1190&---&---&---\\
\hline 40&2012/03/05 04:00:05&  --- &---&---&----&2012/03/07 03:28&2012/03/07 20:50&2012/03/08 11:41\\
\hline 41&2012/03/07 00:24:06&E36,N33& 47&140&2012&2012/03/08 10:54&2012/03/09 05:19&2012/03/11 08:03\\
\hline 42&2012/03/09 04:26:09&W01,N06&  6& 73&1188&---&---&---\\
\hline 43&2012/03/10 18:12:06&W16,N18& 23&107&1271&2012/03/12 08:17&2012/03/12 21:41&2012/03/15 08:42\\
\hline 44&2012/03/13 17:36:05&W37,N33& 47&104&1525&2012/03/15 12:05&2012/03/16 00:51&2012/03/16 12:09\\
\hline 45&2012/04/05 21:25:07&  --- &---&---&----&---&---&---\\
\hline 46&2012/04/09 12:36:07&W40,N12& 41& 94& 892&---&---&---\\
\hline 47&2012/05/12 00:00:05&E25,S10& 26& 65& 939&---&---&---\\
\hline 48&2012/05/17 01:48:05&  --- &---&---&----&---&---&---\\
\hline 
\end{tabular} 
\label{list}
\end{table*}

\begin{figure}
\center
 \noindent\includegraphics[width=0.8\hsize]{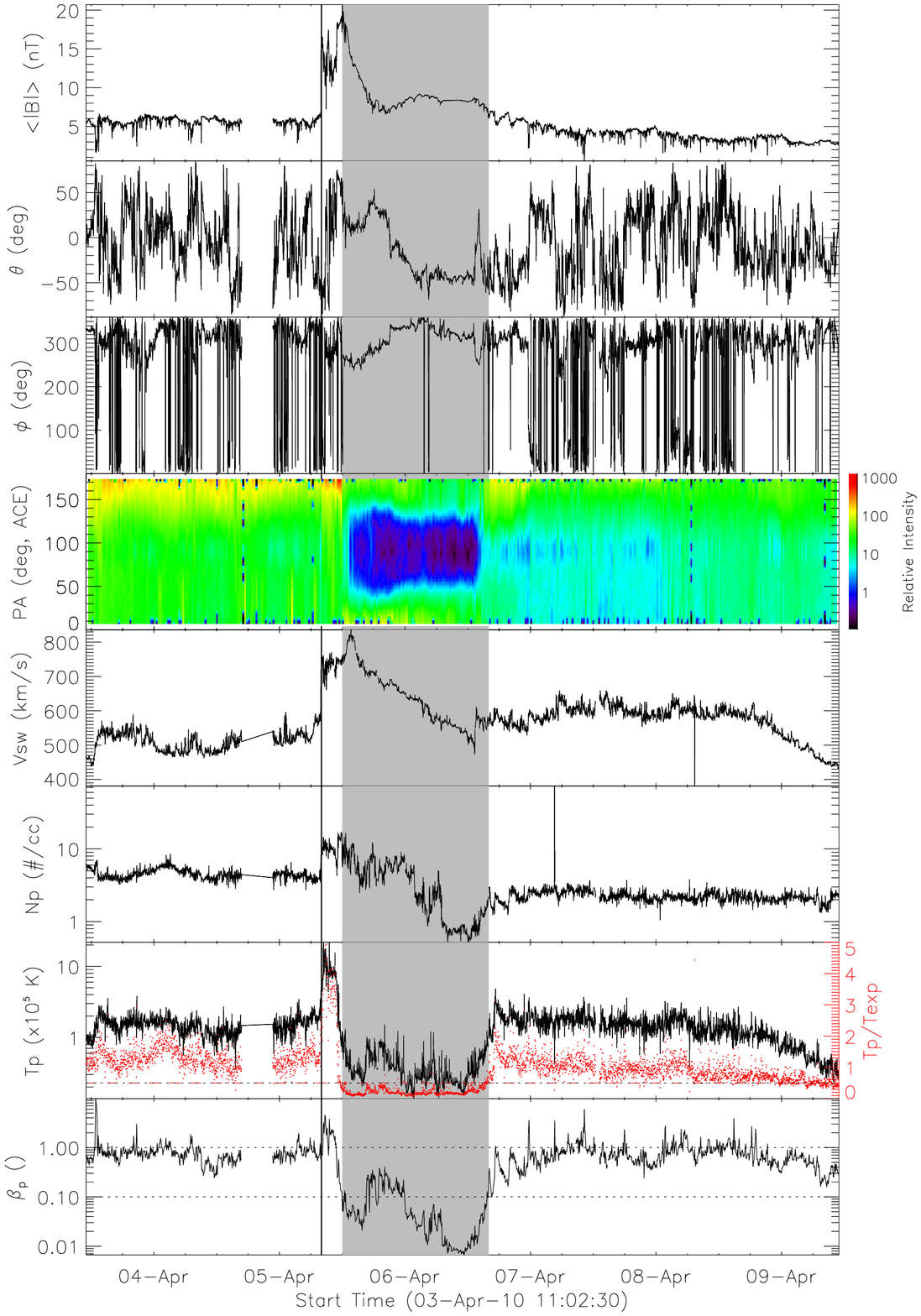}
 \caption{The WIND and ACE observations from 2010 April 3 11:00UT to 2010 April 9 11:00UT. From top to the bottom, they are the magnetic field
 strength ($|B|$) from WIND, the elevation ($\theta$) and azimuthal ($\phi$) of
 field direction based on WIND observations, the ditribution of the electron intensity at different position angle from ACE, solar wind speed ($V_{SW}$)  from WIND, proton density ($N_p$) from WIND, proton temperature ($T_p$) from WIND and the ratio of proton thermal pressure to magnetic pressure ($\beta_p$) calculated based on the WIND observations.}
\label{in_situ}
 \end{figure}

\normalsize
The interplanetary magnetic field and the solar wind plasma observations from WIND and ACE satellites are used to examine whether an interplanetary CME (ICME) was recorded near the Earth the six days following the launch of the CME with an Earthward potential. Previous works used different criteria to identify the ICME\citep[e. g. ][and reference there in]{Wang:2002ki,Wang:2004fp,Jian2006}.
In this paper, the following characteristics are used in the investigation: (1) enhanced magnetic field intensity, (2) smoothly changing field direction, (3) relatively low proton temperature, (4) low proton plasma beta, and (5) bidirectional streaming of electrons. An ICME structure is recognized when it fits at least three of the criteria listed above. The detailed observations of the identified ICME are provided in our online website (http://space.ustc.edu.cn/dreams/fhcmes/). 
Figure \ref{in_situ} shows the in situ observations from 2010 April 3 11:00UT to 2010 April 9 11:00UT, in the next six days after the launch of the CME occurred at 2010 April 3 10:33:58UT. For this event, an obvious ICME was recorded from 2010 April 5 12:00UT to 2010 April 6 16:00UT by WIND and ACE, which is indicated by the gray region in the Figure \ref{in_situ}. This ICME is treated as the possible interplanetary counterpart of the 2010 April 3 10:33:58UT CME. In addition, a shock ahead of this ICME impacted the Earth at approximately 2010 April 5 08:00UT.
 
 \begin{figure}
\center
 \noindent\includegraphics[width=\hsize]{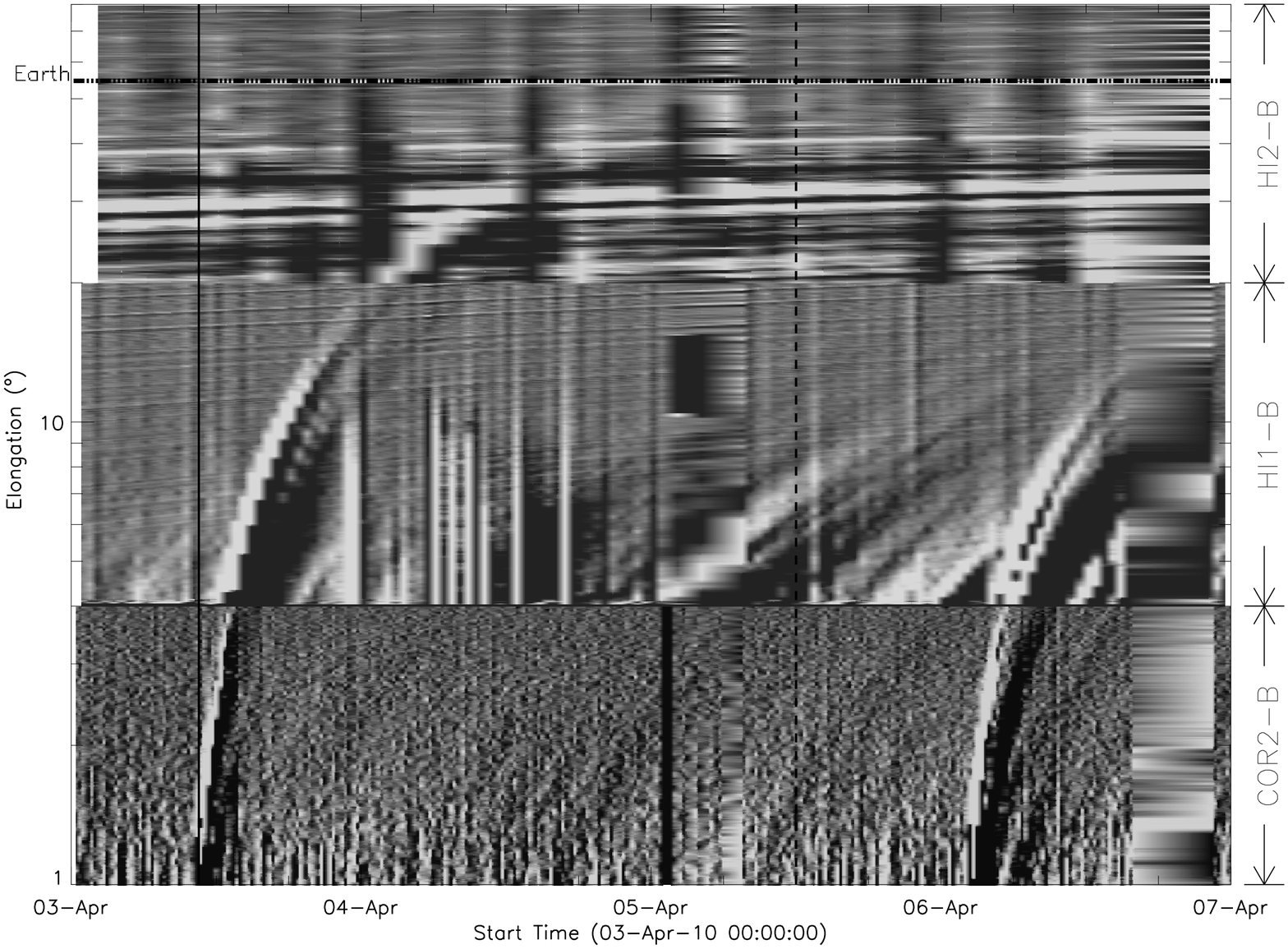}
 \caption{ The time-elonagation (J-map) from 2010 April 3 to 2010 April 7. The solid vertical line shows the time of the CME butst. The dashed line shows the time of the ICME recorded at 1AU. The horizotal dashed-dotted line shows the elongation angle of the Earth.}
\label{jmap}
 \end{figure}

If there is at least one ICME recorded in the next six days after a CME was launched, the Time-Elongation Angle maps (J-maps) \citep[e. g. ][]{1999JGR...10424739S,Davis2009} are used to perform further verification of the association between the ICMEs recorded near the Earth and the FFHCMEs near the Sun.  
A 64-pixel-wide slice is placed along the ecliptical plane in the running-difference images from COR2, HI1 and HI2 on board STEREO, and the slices adopted at different times are stacked to obtain the J-map. A 64-pixel corresponds to $\sim$0.95 solar radius for COR2 and $\sim$1.25$^\circ$ and $\sim$4.38$^\circ$
of the elonagtion angle in the HI1 and HI2 field of view respectively. 
Figure \ref{jmap} shows the J-map for the 2010 April 3 10:33:58UT CME. As seen in Figure \ref{jmap}, after the CME take-off (the vertical solid line), a black-white track that corresponds to the front of this CME extended to the region with an elongation angle of $\sim$60$^\circ$. At the time that the ICME was observed at 1 AU (indicated by the vertical dashed line), the front of this CME also reached a location near the Earth (shown by the horizontal dashed-dotted line). Thus, the ICME observed near the Earth from 2010 April 5 12:00UT to 2010 April 6 16:00UT was well associated with the 2010 April 3 10:33:58UT CME. In this manner, we found that a total of 27 of the 48 FFHCMEs considered hit the Earth. The 7th column of Table \ref{list} shows the arrival time of the shock driven by the CME. The 8th and 9th columns present the beginning and ending times of the associated ICMEs. The `---' symbol denotes that no ICME or shock was associated with this FFHCME.

\section{Which CMEs arrived at the Earth?}

 \begin{figure}
\center
 \noindent\includegraphics[width=0.8\hsize]{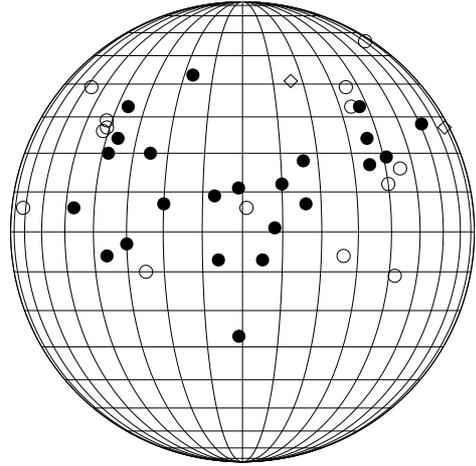}
 \caption{ The distribution of the propagation directions of these FFHCMEs. The solid dots show the Earth-encountered FHCME events 
while the open circles show the events which did not hit the Earth. The diamonds show the events in which only shocks driven by the CMEs hit the Earth.}
\label{sphere}
 \end{figure}

Figure \ref{sphere} shows the distribution of the propagation directions of all 39 FFHCMEs whose de-projected parameters were well established. 
It should be noted that, in the following analysis, only these 39 FFHCMEs were used.
Those propagation directions were distributed in a large range, from E73$^\circ$ to W71$^\circ$ of longitude. According to their propagation longitudes, all 39 FFHCMEs can be classified into 20 eastern events and 19 western events. However, the propagation directions distribution in the north and south sides exhibit an obvious asymmetry. There are a total of 30 events propagating in the northern heliosphere and only nine events in the southern heliosphere. This asymmetry might be caused by the fact that the northern hemisphere is more active than the southern hemisphere in the ascending phase of the 24th solar cycle\citep[e.g.][]{Svalgaard:2012tt}.

In the 39 events whose de-projected parameters have been obtained, 59\% (23) of them hit the Earth. The solid dots in Figure \ref{sphere} show the events that arrived at the Earth. The figure shows that their propagation longitudes are distributed in a range of [E47$^\circ$, W61$^\circ$], which is more narrow than the longitudinal distributions of all FFHCMEs. Note that the most eastern of the Earth-encountered events propagated at E47$^\circ$, and the most western event came from W61$^\circ$. This observation is consistent with the previous result that it is difficult for the east limb CMEs to hit the Earth\citep{Wang:2002ki,2003ApJ...582..520Z}.
In addition, the longitude range of most (21/23) of the Earth-encountered FFHCMEs is [E40$^\circ$,W40$^\circ$]. Meanwhile, for the FFHCMEs whose propagating longitudes located in this region, approximately 72\% of them arrived at the Earth. This result suggests that the central events with propagation longitudes in the range of [E40$^\circ$, W40$^\circ$] are more likely to hit the Earth.

\begin{figure}
\center
 \noindent\includegraphics[width=\hsize]{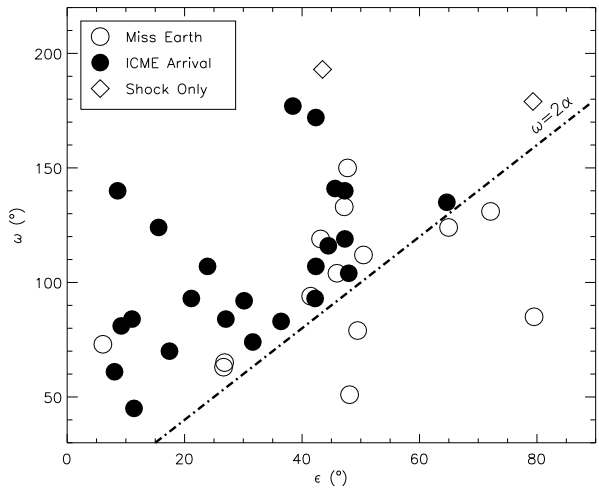}
 \caption{The angular width $\omega$ varied with $\epsilon$ for the FFHCMEs. 
The solid dots show the Earth-encountered CMEs while the open circles show the not Earth-encountered CMEs. The diamonds show the events in which only shocks driven by the CMEs hit the Earth.}
\label{wdf}
 \end{figure}

The deviation angle $\epsilon$, which is defined as the angle between the propagation direction and the Sun-Earth line, is used to discriminate among the possible Earth-encountered CMEs. 
From Figure \ref{wdf}, 24 FFHCMEs propagated with $\epsilon \le 45^\circ$ and 75\% (18) of them hit the Earth. 
Meanwhile,  for the 13 FFHCMEs with $\epsilon \le 30^\circ$, 77\% (10) of them arrived at the Earth.
For comparison, only one main body or the flux rope structure of five limb FFHCMEs with $\epsilon \ge 60^\circ$ hit the Earth. It should be noted that, one shock driven by another limb CME hit the Earth. But, in this work, we mainly discussed whether the main body or the flux rope like structure of the CME hit the Earth.  This observation confirms the previous result that the central CMEs can hit the Earth with higher possibility. In addition, five events with $\epsilon > 45^\circ$ arrived at the Earth. Upon checking their de-projection parameters, we find that all of these events are wide events and that their minimum angular width is 103$^\circ$. This observation indicates that the limb CMEs can also impact the Earth if they are wide.

 \begin{figure}
\center
 \noindent\includegraphics[width=0.8\hsize]{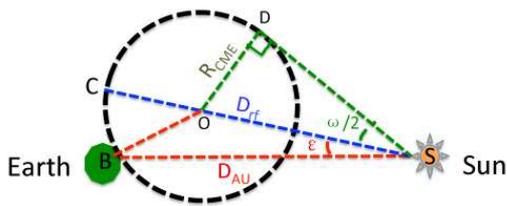}
 \caption{The sketch map of the CME's self-similar expansion model.
 }\label{sketch}
 \end{figure}

Simply by assuming that a CME moves as a self-similar expansion ball  \citep[e.g.][]{Davies2012,Mostl2012a} as shown in Figure \ref{sketch}, 
one can expect that the CME can hit the Earth when its angular width is larger than twice the deviation angle, i.e., $\omega > 2\epsilon$. The dotted-dashed line in Figure 5 indicates $\omega = 2\epsilon$. Based on the previous analysis, only CMEs located in the region above this line could hit the Earth. The observations that all of the Earth-encountered FFHCMEs are located in the upper region confirm the above conclusion. In addition, a large fraction (74\%, 25/34) of FFHCMEs that fit the condition of $\omega>2\epsilon$ hit the Earth. For comparison, all four events under the dotted-dashed line did not hit the Earth. This result indicates that the $\omega > 2\epsilon$ can be a useful criterion to forecast whether a CME would hit the Earth; therefore, the angular width is another important parameter in the space weather forecasting model.

Note that some CME events that were launched from regions close to the solar-disc center or fit the criterion of $\omega > 2\epsilon$ did not hit the Earth. One possible reason is that those CMEs may be deflected during their interaction with other CMEs\citep[e.g.][]{Xiong2009,Wang:2011cm,Lugaz2012a,Shen2012a}. 
\citet{Lugaz2012a}  studied the interaction between the CMEs occurred on 2010 May 23 and 2010 May 24, which are the No 5 and No 6 events, respectively, in our list. They found that the Earth-direct CME on 2010 May 24 had missed the Earth, which was a result of the interaction with the 2010 May 23 CME. In our investigation, six events with $\epsilon < 45^\circ$ and $\omega > 2\epsilon$ did not hit the Earth. By carefully checking the heliospheric images from STEREO HI1 and HI2, we found that two events might be affected by their interaction with other CMEs: the 2012 May 24 event, which was studied by \citet{Lugaz2012a}, and the 2012 March 9 event.
Why did the CMEs of the other four events not hit the Earth? A possible reason is that the CMEs might be deflected during their propagation in the corona
\citep[e.g.][]{MacQueen1986,Gopalswamy2000a,Cremades:2004cz,Kilpua2009a,Gui:2011dw,Shen2011,Zhou2013} as well as in the interplanetary space\citep[e.g.][]{Wang:2004fp,Wang2006c,Wang2013b}.

\section{When did CMEs arrive at the Earth?}

 \begin{figure}
\center
 \noindent\includegraphics[width=\hsize]{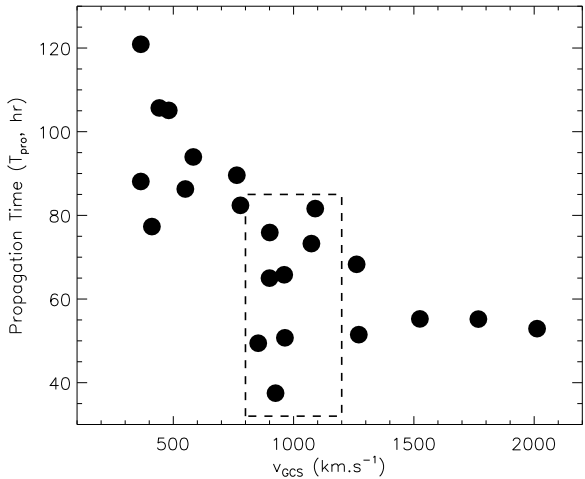}
 \caption{The propagation times varied with the $v_{GCS}$ of the Earth-encountered FFHCMEs. The rectangle include the eight events with similar velocity of 1000$\pm$200 km.s$^{-1}$.}
\label{tpro}
 \end{figure}

Figure \ref{tpro} shows that the transit times of the CMEs from the Sun to 1 AU varied with the de-projection velocities of the CMEs. 
In Figure \ref{tpro}, the propagation time exhibits an anti-correlation with the de-projected velocities, but the dispersion is large. 
For the CMEs with similar velocities, the difference of the transit time can be up to tens of hours. For example, there are eight events whose velocities are $\approx$1000$km.s^{-1}$ (from 800$km.s^{-1}$ to 1200$km.s^{-1}$). The transit time of these events from the Sun to 1 AU varied from 37 hours for the 2011 August 3 event to 82 hours for the 2012 January 19 event. The difference between the propagation times is approximately 2 days.
 
Why do these CMEs with similar velocities have quite a different transit time from the Sun to 1 AU? It is probably because the CMEs have a circular-like front, and it is not always true that the leading front of a CME encounters the Earth. This effect means that in the case of `non-central' impact, the CME forehead reaches a distance larger than 1 AU at the time when the CME arrival is recorded in the in situ observations near the Earth. A well-investigated case can be found in the most recent work by \citet{Wang2013b}, 
in which a CME occurred on 2008 September 13 passed through both the WIND and STEREO-B spacecraft at 1 AU. The arrival time of this CME at STEREO-B was approximately two days later than that at WIND. Again, we assume that the CME is a self-similar expansion ball radially propagating along the direction with a deviation angle of $\epsilon$. When the observatory at 1 AU detected this CME, the real propagation distance of the CME tip along its propagation direction ($D_{rf}$, the length of SC in Figure \ref{sketch}) was obviously larger than 1 AU ($D_{AU}$). The difference between the $D_{rf}$ and 1 AU, $\Delta D=D_{rf}-1AU$, depends on the angular width of the CME ($\omega$) and the propagation direction $\epsilon$. The $D_{rf}$ could be obtained by:
\begin{linenomath*}
\begin{eqnarray}
D_{rf}=R_{CME}+\frac{R_{CME}}{sin(\frac{\omega}{2})}\label{eq_1}
\end{eqnarray}
\end{linenomath*}
in which $R_{CME}$ is the radius of this CME, which can be calculated from the equation of: 
\begin{linenomath*}
\begin{eqnarray}
(\frac{R_{CME}}{sin(\frac{\omega}{2})} )^2+D_{AU}^2-2D_{AU}\frac{R_{CME}}{sin(\frac{\omega}{2})}cos\epsilon=R_{CME}^2\label{eq_2}
\end{eqnarray}
\end{linenomath*}

\begin{figure}
\center
 \noindent\includegraphics[width=0.8\hsize]{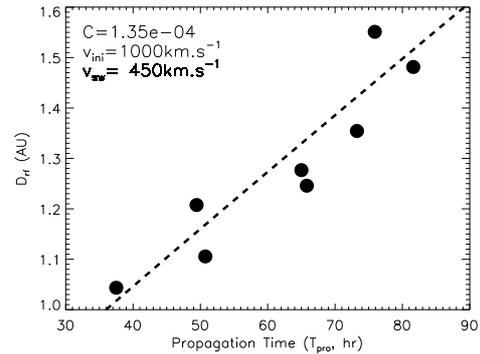}
 \caption{The CME's real propagation distances $D_{rf}$ varied with propagation times for the Earth-encountered FFHCMEs with $800km.s^{-1}\le v_{gcs} \le 1200km.s^{-1}$. 
The real proapagtion distances $D_{rf}$ are calculated based on the SSE model and the propagation time is obtained from the real obsrvations.}\label{realdis}
 \end{figure}

From Equations 1 and 2, the real propagation distance $D_{rf}$ can be determined if the angular width $\omega$ and deviation angle $\epsilon$ are all known. To improve the comparison, we choose eight events with similar velocities in the range of $1000 \pm 200km.s^{-1}$ for the further study. 
Figure \ref{realdis} shows that calculated real propagation distance $D_{rf}$ varied with the observed propagation time of these eight events based on the de-projected parameters obtained in paper I. 
Additionally, the $D_{rf}$ of these events varied over a large range, from 1.04 AU to 1.53 AU. 
As shown in the previous analysis, $D_{rf}$ could be treated as the real propagation distance of these CMEs along their propagation direction when they eventually arrived at 1 AU. 
As shown in Figure \ref{realdis}, the propagation time and the real propagating distance of these events have obvious positive correlation. 
This result indicates that the different transit time of CMEs with similar velocities might be caused primarily by the different part of the circular-like CME front arriving at 1 AU. 
From equations \ref{eq_1} and \ref{eq_2}, the propagation distance of the CME tip is related to the angular width and the propagation direction. 
Thus, the true angular width and the propagation direction are all important parameters in the CME arrival time forecasting as well as the CME's velocity and the background solar wind speed
\citep[e.g.][]{Vrsnak2007d,Temmer2011}.

The propagation of CMEs in the interplanetary space can be described by an aerodynamic drag model\citep[e.g.][and reference therein]{Chen1996,Maloney2010,Vrsnak2012,Lugaz:2012es}.
Here, we use a simplified equation of the aerodynamic drag model from \citet{Maloney2010} as:
\begin{linenomath*}
\begin{eqnarray}
\frac{dv_{CME}}{dr}=-Cr^{-1/2}(v_{CME}-v_{SWE})^2\label{eq_df}
\end{eqnarray}
\end{linenomath*}
in which C is a constant number. The dashed line in Figure \ref{realdis} shows the result of the aerodynamic drag model by assuming the initiation speed of a CME is 1000$km.s^{-1}$, the solar wind speed $v_{SWE}$ is 450 $km.s^{-1}$. The value $C=1.35 \times 10^4$ is obtained from a fitting process. In this figure, almost all of the points are close to the dashed line. Therefore, the propagation process of these CMEs could be well described by the aerodynamic drag model. Thus, the self-similar expansion model combined with the aerodynamic drag model might be a powerful tool to forecast the Earth arrival times of CMEs.

\section{Conclusion and Discussion}

In this work, we studied whether and when the front-side full halo CMEs that occurred from 2007 March 1 to 2012 May 31, hit the Earth. The de-projected parameters of those CMEs were obtained in our previous work\citep{Shen2013d}.  
The in situ observations combined with the SECCHI/COR2 SECCHI/HI-1 and SECCHI/HI-2 observations from STEREO were used to verify whether these FFHCMEs hit the Earth. We conclude the following:

\begin{enumerate}
\item Approximately 59\% of the FFHCMEs studied in this work arrived at the Earth. The central CMEs, which propagated in the longitudinal range [E40$^\circ$, W40$^\circ$] or $ \epsilon \le 45^\circ$, can arrive at the Earth with higher probability. 

\item The FFHCMEs with an angular width $\omega$ of more than twice the deviation angle $\epsilon$ can hit the Earth. All of the Earth-encountered events fit the criterion $\omega > 2\epsilon$, and 74\% of the FFHCMEs events that fit the criterion of $\omega > 2\epsilon$ hit the Earth. Thus, the simple criterion ($\omega > 2\epsilon$) might be a useful tool to forecast whether a CME will hit the Earth. 

\item The propagation times exhibit an overall anti-correlation with the de-projected velocities. The self-similar expansion model can be used to adequately explain the different transit time of the CMEs from the Sun to 1 AU with similar velocities. Furthermore, we suggest that the self-similar expansion model combined with the aerodynamic model is a simple and useful tool to forecast the arrival time of CME. 
\end{enumerate}

\begin{figure}
\center
 \noindent\includegraphics[width=0.8\hsize]{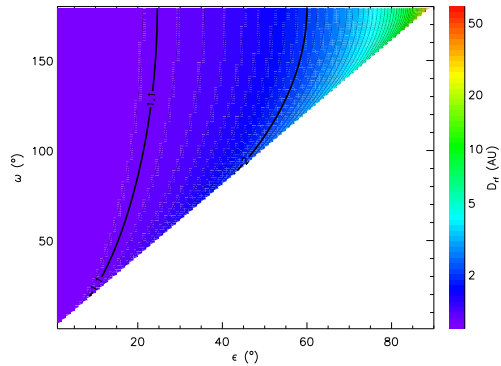}
 \caption{The dependence of $D_{rf}$ on different combinations of the angular width $\omega$ and deviation angle $\epsilon$.}\label{sse}
 \end{figure}

\begin{figure}
\center
 \noindent\includegraphics[width=0.8\hsize]{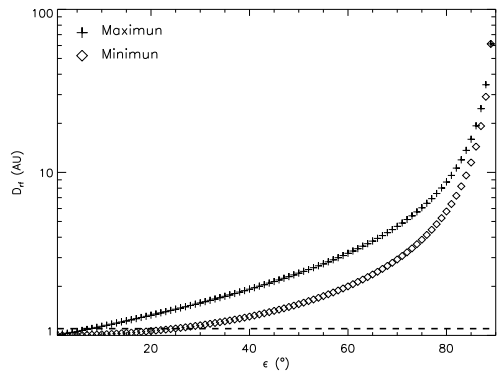}
 \caption{The maximum and minimum values of $D_{rf}$ as a function of the deviation angle $\epsilon$. The horizontal dashed line shows the $D_r=1.1$ AU.}\label{maxsse}
 \end{figure}

Based on the previous analysis, we found that the CME's real propagating distance, $D_{rf}$, is determined by its angular width and the propagation direction. 
The propagating distance of the CME tip might be larger than 1 AU when the flank of the CME hit the Earth. 
To evaluate the influence of this effect, the values of $D_{rf}$ with different angular widths $\omega$ and deviation angles $\epsilon$ were calculated. 
Assuming the deviation angle $\epsilon$ varied from 1$^\circ$ to 90$^\circ$ and the angular width $\omega$ varied from 1$^\circ$ to 180$^\circ$, 
Figure \ref{sse} shows the distribution of the $D_{rf}$ values for the different cases. 
Note that the CMEs in the lower right half of the plot ($\omega < 2\epsilon$, white regions in Figure \ref{sse}) can not hit the Earth based on the SSE model. 
In other cases with $\omega > 2\epsilon$, $D_{rf}$ varied over a wide range, from $\approx$1 AU to more than 60 AU. 
Considering that a 10\% uncertainty in the forecasting of the CME arrival time is acceptable, 
the influence of the different propagation direction and angular width on the travel time of a CME should be carefully considered in the cases with $D_{rf} >$ 1.1 AU. 
In a large fraction (60\%) of the cases we discussed, the $D_{rf}$ are larger than 1.1 AU, and those parameters must be considered. 
Inspecting Figure \ref{sse}, one finds that at a given fixed value of $\epsilon$, $D_{rf}$ increases as the angular width decreases. 
Thus, the $D_{rf}$ for narrower CMEs should be calculated first to forecast the arrival time of the CMEs.
In addition, $D_{rf}$ varies greatly with the change of the deviation angle, as shown in Figure \ref{sse}.
Larger $\epsilon$ values correspond to larger $D_{rf}$ values. 
Figure \ref{maxsse} shows the maximum and minimum values of $D_{rf}$ as a function of the deviation angle $\epsilon$ by assuming that the CME angular width varies from 1$^\circ$ to 180$^\circ$. It is found that both values increase with increasing $\epsilon$. Particularly, when $\epsilon > 25^\circ$, the values of $D_{rf}$ are always larger than 1.1 AU.
Thus, combined with the previous results, we suggest that for narrow CMEs or the CME propagated with $\epsilon > 25^\circ$, the influence of the propagation direction and angular width on the CME's transit distance is large, and the $D_{rf}$ must be carefully calculated in the CME arrival time forecasting.

\acknowledgments{We appreciate using of the  CME catalog, the data
from SECCHI instruments on STEREO and LASCO on SOHO. 
The CME catalog is generated and maintained at the CDAW Data Center by NASA 
and the Catholic University of America in cooperation with the Naval
Research Laboratory. STEREO is the third mission in NASA Solar
Terrestrial Probes program, and SOHO is a mission of international
cooperation between ESA and NASA.
We also thank the NSSDC at Goddard Space Flight Center/NASA for providing Wind and ACE data. 
The data for this paper are available at the official website of SOHO, STEREO, WIND and ACE satellites. 
The dataset we used are SOHO LASCO/C2, SOHO LASCO/C3, STEREO COR2, STEREO HI1, STEREO HI2, WIND MFI, WIND SWE and ACE SWEPAM.
This work is supported by the Chinese Academy of Sciences (KZZD-EW-01), grants from the 973
key project 2011CB811403, NSFC 41131065, 41274173, 40874075, 
41121003, and 41304145, CAS the 100-talent program, KZCX2-YW-QN511 and startup
fund, and MOEC 20113402110001, the fundamental research funds for
the central universities (WK2080000031) and the Strategic Priority Research Program on Space Science, the Chinese Academy of Sciences
(Grant No. XDA04060801).}


\end{document}